# An age-of-allele test of neutrality for transposable element insertions.

Justin P. Blumenstiel[*], Xi Chen[*], Miaomiao He[§] and Casey M. Bergman[§]

[*] Department of Ecology and Evolutionary Biology, University of Kansas, Lawrence, KS, 66049

[§] Faculty of Life Sciences, University of Manchester, Manchester, UK, M21 0RG

7   **Corresponding author:**
8   Justin P. Blumenstiel
9   Department of Ecology & Evolutionary Biology
10  University of Kansas
11  1200 Sunnyside Ave.
12  Lawrence, KS 66049
13
14  Tel: 785-864-3915
15  Email: jblumens@ku.edu
16

**Abstract**

How natural selection acts to limit the proliferation of transposable elements (TEs) in genomes has been of interest to evolutionary biologists for many years. To describe TE dynamics in populations, previous studies have used models of transposition-selection equilibrium that assume a constant rate of transposition. However, since TE invasions are known to happen in bursts through time, this assumption may not be reasonable. Here we propose a test of neutrality for TE insertions that does not rely on the assumption of a constant transposition rate. We consider the case of TE insertions that have been ascertained from a single haploid reference genome sequence. By conditioning on the age of an individual TE insertion allele (inferred by the number of unique substitutions that have occurred within the particular TE sequence since insertion), we determine the probability distribution of the insertion allele frequency in a population sample under neutrality. Taking models of varying population size into account, we then evaluate predictions of our model against allele frequency data from 190 retrotransposon insertions sampled from North American and African populations of *Drosophila melanogaster*. Using this non-equilibrium neutral model, we are able to explain about 80% of the variance in TE insertion allele frequencies based on age alone. Controlling for both non-equilibrium dynamics of transposition and host demography, we provide evidence for negative selection acting against most TEs as well as for positive selection acting on a small subset of TEs. Our work establishes a new framework for the analysis of the evolutionary forces governing large insertion mutations like TEs, gene duplications or other copy number variants.

# Introduction

Natural selection against transposable element (TE) insertions is considered to be one of the primary forces preventing their proliferation in populations. The action of negative selection against these genetic parasites is thought to come in three predominant forms: selection against insertions in functional regions (CHARLESWORTH and LANGLEY 1989), chromosomal abnormalities arising from ectopic recombination (MONTGOMERY et al. 1987; LANGLEY et al. 1988), and costs associated with the transposition process itself (NUZHDIN et al. 1996). Understanding the relative importance of each of these forces has been of substantial interest for many years (CHARLESWORTH and LANGLEY 1989; CHARLESWORTH et al. 1994; NUZHDIN 1999; LEE and LANGLEY 2010). To understand the nature of selection acting on TEs, a common practice is to measure the allele frequency distribution of TE insertions within natural populations (MONTGOMERY et al. 1987; BIEMONT et al. 1994; PETROV et al. 2003; YANG and NUZHDIN 2003; GONZALEZ et al. 2008; PETROV et al. 2011; KOFLER et al. 2012). These studies have found that TE insertion alleles segregate at low allele frequencies in *D. melanogaster*, and this observation has been used to support the idea that negative selection acts to prevent TE insertions from increasing in frequency in populations (CHARLESWORTH and LANGLEY 1989).

A limitation of previous studies on the dynamics of TE evolution is that the frequency distribution under different models of selection is typically evaluated under the assumption of transposition-selection balance within the population (CHARLESWORTH and LANGLEY 1989; PETROV et al. 2003; LOCKTON et al. 2008; GONZALEZ et al. 2009; LEE and LANGLEY 2010). A crucial assumption of models that posit transposition-selection balance is that the transposition process can be modeled as a constant rate over time. This is often unlikely to be the case, as episodes of transposition are known to occur in bursts. For example, the *P*-element invaded and proliferated in *D. melanogaster* only within the past several decades (KIDWELL 1983; DANIELS *et al.* 1990). Likewise, analysis of genome sequences has demonstrated waves of transposition for a number of other TE families (SANMIGUEL *et al.* 1998; LANDER *et al.* 2001; DE LA CHAUX and WAGNER 2009; LU *et al.* 2012). In cases of recent transposition bursts, insertion allele frequencies will not be at equilibrium because there will not have been sufficient time to drift to moderate or high allele frequencies, even under strict neutrality. Therefore recent insertion alone may explain the pattern of low allele frequencies for TE insertions observed in natural populations of *D. melanogaster* (BERGMAN and BENSASSON 2007). Alternatively, negative selection may explain the pattern since equilibrium can be achieved quickly when TEs are harmful. To distinguish among these possibilities, it would be beneficial to relax the assumption of transposition-selection balance in models of TE evolution. We develop such an approach here. To relax equilibrium assumptions we ask: are TE insertion allele frequencies consistent with neutrality, conditional on the inferred time that has elapsed since insertion? If so, then one may conclude that genetic drift and demography are the major factors shaping the evolution of TE insertion allele frequencies. However, If TE insertions are observed at a lower frequency than predicted based on their age, we may infer that negative selection is limiting their increase. Alternately, if a TE insertion is at a higher frequency than expected based on its age, we may infer the action of positive selection acting on that locus.

Critical to this approach is being able to estimate the time that has elapsed since origination of the insertion allele. For most mutations, information about allele age is provided solely by the frequency of the allele itself or in the amount of linked variation (SLATKIN 2000). Under neutrality, a low frequency allele is on average younger than a high frequency allele (KIMURA and OHTA 1973) and alleles with low levels of linked variation and greater haplotype structure tend to be younger because there has not been sufficient time to accumulate mutations or undergo recombination (SLATKIN 2000).

For large insertions like TEs, an additional source of age information can be obtained from the insertion sequence itself. Specifically, the age of a TE insertion ascertained from a single genome can be inferred by estimating the number of unique substitutions that have accumulated in the TE sequence since its insertion, relative to the entire transposing lineage. After insertion, most TE sequences evolve under an unconstrained, pseudogene-like mode of evolution (PETROV *et al.* 1996). Thus, by determining the number of nucleotide differences between the actively transposing lineages and a particular TE insertion, one can estimate the age of that particular insertion event under the standard assumptions of a molecular clock. Dating the age of TE insertions (in terms of nucleotide substitutions on their terminal branches) has been proven instrumental in determining spontaneous rates of insertion and deletion in *Drosophila* where classical pseudogenes are relatively rare (PETROV *et al.* 1996). Information about the age since insertion has also previously proven useful in understanding the dynamics of TE invasion in the history of a species (BERGMAN and BENSASSON 2007).

Here we use results from coalescent theory to determine the neutral probability distribution of allele frequency for a neutral TE insertion identified in a reference genome, conditional on its estimated time since insertion. This method is particularly suitable for genotyping or resequencing studies in which TEs identified in a well-assembled genome are subsequently assayed for their allele frequency in populations (BLUMENSTIEL *et al.* 2002; PETROV *et al.* 2003; FRANCHINI *et al.* 2004; NEAFSEY *et al.* 2004; LIPATOV *et al.* 2005; GONZALEZ *et al.* 2008; PETROV *et al.* 2011). Since the age of an insertion allele cannot be exactly determined, we incorporate uncertainty in age estimates into our approach by integrating over the Bayesian posterior distribution of time since insertion. Our approach allows one to test whether TE insertion frequencies are as expected under neutrality, without assuming constancy of transposition rate or constant host population size. We apply this method to a sample of 190 retrotransposon insertions in *D. melanogaster* that have previously been shown to undergo the pseudogene-like mode of sequence evolution (BERGMAN and BENSASSON 2007). Using published demographic scenarios of population history in *D. melanogaster* as examplars, we demonstrate that a neutral model that takes age of insertion into account can explain more than 80% of the variation in TE insertion frequencies. In addition, we show how conditioning on time since insertion enables the detection of negative and positive selection acting on TEs without assuming equilibrium TE and host dynamics.

## Materials and Methods

*Estimating the age of a TE insertion ascertained from a reference genome*

To estimate time of TE insertion we count the number of substitutions ($s$) that are unique to a particular TE insertion relative to all other sequenced paralogous copies of a TE family residing in a single reference genome (Fig 1A). We note that our estimate of age is not the time to the most recent common ancestor (MRCA) based on intra-allelic variation within orthologous copies of a particular insertion allele (see Fig 1B). We discount substitutions that are shared among paralogous copies because these represent differences that occurred prior to the origin of distinct, actively transposing lineages of the same family. Assuming that a newly inserted TE is not co-opted for some function by the host, unique substitutions within a TE sequence accumulate under an unconstrained, pseudogene-like mode of evolution and these can serve as a measure of time since insertion. A lack of constraint on substitutions after insertion can be demonstrated by generating multiple alignments of paralogous TE copies of a family within a single reference genome and identifying substitutions that occur on active TE lineages (shared among copies) versus those that occur within individual TE insertions (unique to single copies). Previous work has shown that shared substitutions are only abundant at third positions within codons, consistent with selection to maintain a functional amino acid sequence, whereas unique substitutions do not show this pattern (PETROV *et al.* 1996; BERGMAN and BENSASSON 2007). A limitation of this method is that TEs that have inserted in the very recent past will all have zero unique substitutions, making it difficult to precisely determine how old they are.

*The probability of i copies in a sample of n alleles, conditional on the age of an insertion sequence*

To relax the assumption of transposition-selection balance in a test of neutrality for TEs, we seek the probability distribution of neutral TE insertion allele frequency conditional only on time of allele origination and host population history. In this way, we free ourselves from the assumption of a constant transposition rate because we focus only on the individual frequency of a retrotransposon insertion allele that cannot excise. This is in contrast to approaches that compare the entire distribution of allele frequencies from multiple TE insertions with an equilibrium distribution generated assuming constant transposition rate.

Here we determine the probability that a neutral TE insertion allele ascertained from a haploid genome will be present in $i$ copies in a sample of $n$ alleles, conditional on the time since insertion (Figure 1B). An example of this approach was previously used to discriminate TEs based on age that are expected to be polymorphic rather than fixed in pufferfish (NEAFSEY et al. 2004). This probability is conditional on 1) the number of sample ancestors present at time $t$ of insertion, 2) the probability that a lineage which received the insertion at time $t$ is represented in $i$ descendants within $n$ sampled alleles and 3) ascertainment of the TE insertion from a single haploid genome (which specifies the ancestor at time $t$).

The probability than $n$ sampled alleles have $j$ ancestors at time $t$ is given by:

$$P(j|t,n) = \sum_{k=j}^{n} \rho_k(t) \frac{(2k-1)(-1)^{k-j} j_{(k-1)} n_{[k]}}{j!(k-j)! n_{(k)}}, \quad 2 \leq j \leq n$$

$$P(j|t,n) = 1 - \sum_{k=2}^{n} \rho_k(t) \frac{(2k-1)(-1)^k n_{[k]}}{i_{(k)}} \quad j=1$$

(1)

where $\rho_k(t) = \exp\{-k(k-1)t/2\}$, $a_{(k)} = a(a+1)...(a+k-1)$, $a_{[k]} = a(a-1)...(a-k+1)$ and $t$ is in units of $N_e$ (the effective population size) generations under a haploid model or $2N_e$ generations under a diploid, two sex model (TAVARE 1984; MOHLE 1998). In this treatment, we consider scenarios of varying population size that have been proposed by others for *Drosophila* (LI and STEPHAN 2006; DUCHEN *et al.* 2013). To achieve this, *t* in equation 1 can be rescaled as a function of $2N_e$ at appropriate times (GRIFFITHS and TAVARE 1998). For example, going backwards in time, a halving of the population size at a given time would lead to *t* being scaled in $2N_e/2$ generations at the point and further backwards.

Conditional on *j* ancestors at time *t*, the probability that a single non-specified ancestor that received an insertion is represented by *i* copies in a sample of size *n* under random sampling is a combinatorial probability given by:

$$P(i \mid j,n) = \frac{(j-1)(n-i-1)!(n-j)!}{(n-1)!(n-j-i+1)!} \qquad (2)$$

(SLATKIN 1996; SHERRY *et al.* 1997). Here, *i* ranges from 1 (only the ascertained allele is present) to *n* (fixed in the sample) and we define the probability equal to zero when $i > (n - j) + 1$. When *j* equals 1, we define all the probability to be at $i = n$ and when $j = n$ we define all the probability to be at $i = 1$. Interestingly, and as pointed out by others (FELSENSTEIN 1992; SHERRY *et al.* 1997), when $j=2$, the probability distribution is uniform from *i = 1* to *n-1*. Note that *n* here includes the haploid genome sample from which the insertion was ascertained and the single ancestor of the haploid genome, among *j* ancestors, is not specified. In fact, equation 2 is the probability for any chosen single ancestor out of *j* ancestors, which may not be a reference genome ancestor. Accounting for specification of the reference ancestor is achieved in the correction for ascertainment bias below.

An assumption of this model is that there are no full-length excisions of the TE over this time period. This assumption is valid for RNA-based retrotransposons but make the model less applicable to DNA-based transposons which undergo excisions resulting in descendants of the specified ancestor that subsequently lack the insertion.

Combining equations (1) and (2), the probability of $i$ copies, conditional on $t$ time of insertion and $n$ samples, is given by the probability of $i$ copies conditional on $j$ ancestors, multiplied by the probability of $j$ ancestors conditional on time $t$, summed over all $j$:

$$P(i|t,n) = \sum_{j=1}^{n} P(i|j,n) P(j|t,n) \qquad (3)$$

Equation (3) provides the probability that an insertion that occurred at time $t$ is present in a sample on $n$ alleles, but it does not account for how the allele was discovered. For TE insertions identified in a single reference genome sequence, there is ascertainment bias since insertions that occur at time $t$ and are absent from the reference but present elsewhere in the sample are ignored. In this way, the single ancestor for which the insertion occurred at time $t$ must be specified. To deal with this ascertainment bias, it is necessary to condition on the probability that a TE of a certain specified allele count (designated $i_s$) in a sample $n$ is in the reference genome sequence. The probability of being ascertained in the genome is the frequency in the total sample that includes the genomic reference: $i_s/n$. Therefore, the final probability of $i$ conditional on ascertainment - designated $i_a$ - is given by:

$$P(i_a|t,n) = \frac{\dfrac{i_f}{n} P(i_f|t,n)}{\sum_{i=1}^{n} \dfrac{i}{n} P(i|t,n)} \qquad (4)$$

*Accounting for Error in Age Estimation*

This formulation assumes that the age of the insertion is known absolutely, which is not the case. For a particular insertion, the uncertainty in its age estimate will be a function of the number of substitutions as well as the size of the element. For TE insertions with an equivalent proportion of unique substitutions, larger insertions will provide more accurate age estimates. Therefore, rather than assuming that the time of insertion is known, it is desirable to condition on the probability distribution of the insertion age. By doing so, one can determine the probability distribution of allele frequency in a sample given the size of the TE insertion as well as the number of substitutions that it has received since insertion. Using Bayes' rule and assuming substitutions occur according to a simple Poisson process, the probability distribution of time of insertion $t$ is given by 1) the probability of $s$ substitutions in fragment of length $l$, conditional on a specified $t$ (designated $t_s$), multiplied by the probability of $t_s$, divided by 2) the probability of $s$ substitutions in fragment of length $l$, integrated over all time ($t$):

$$P(t_s | s_l) = \frac{P(s_l | t_s) P(t_s)}{\int_0^\infty P(s_l | t) P(t) dt} \quad (5)$$

In this case the prior probability distribution is $P(t)$ and $P(s_l|t)$ is determined using the Poisson distribution with the parameter $\lambda$:

$$P(s|t) = \frac{(t\lambda)^s e^{-t\lambda}}{s!} \quad (6)$$

Here, the Poisson parameter $\lambda$ is the expected number of mutations per generation in a sequence of length $l$ given a fixed mutation rate per base pair, $u$. We used an empirical Bayes approach in which the distribution of the number of substitutions within all TE insertions ascertained from the reference genome was used to estimate the parameters of the prior distribution of time since insertion - here chosen to either be an exponential or gamma. Exponential/gamma priors were chosen based on their common and analagous use in Bayesian estimation of branch lengths (HUELSENBECK and RONQUIST 2001; YANG and RANNALA 2005; HEATH 2012).

For TEs with zero substitutions since the time of insertion, the maximum likelihood estimate for *t* will approach zero, but such a TE will always be at least slightly older than this. Within a Bayesian framework, longer TE insertions with zero substitutions will be estimated to be younger than smaller insertions that also have zero substitutions since smaller insertions have less power in updating the prior. We also assume that the number of substitutions found in the TE sequence is small enough such that a correction for multiple hits is not necessary. The full probability distribution incorporating uncertainty in age estimates is:

$$P(i \mid s_l, n) = \sum_{j=1}^{n} \left( \int_0^\infty P(j \mid t, n) P(t \mid s_l) dt \right) P(i \mid j, n) \qquad (7)$$

where the integral term in the parentheses is probability of *j* ancestors (equation 1) conditional on time of insertion (equation 5), integrated over all insertion times. The remaining probability (on the right hand side of equation 7) is the probability of *i* alleles in a sample conditional on *j* ancestors (equation 2). The full probability is the probability of *i* alleles, conditional on *j* ancestors, summed over all possible numbers of ancestors. Bias due to ascertainment in equation 7 can be corrected as above using equation 4.

*Estimation of TE insertion allele frequency in D. melanogaster populations*

We selected 190 loci from LTR and non-LTR retrotransposon families whose sequences have been shown previously to evolve under a pseudogene-like mode of molecular evolution in *D. melanogaster* (BERGMAN and BENSASSON 2007) and that also had PCR primer sequences available in the literature (GONZALEZ *et al.* 2008). We did not sample any DNA transposon families, since their ability to transpose through a DNA intermediate violates the assumption that the number of unique substitutions represents its time since integration. Families were chosen on the basis of maximal coverage of loci in an alignment (not family age or size). In total, we sampled 90 LTR and 100 non-LTR elements from the following families (sample sizes in parentheses): *copia* (23), *burdock* (12), *blood* (19), *412* (23), *17.6* (8), *micropia* (2), *rover* (2), *invader2* (1), *BS* (11), *Cr1a* (18), *Doc* (42), *G4* (8), *G5* (4), *Helena* (5), *Juan* (7), *baggins* (2), *jockey2* (2) and *Doc3* (1). Age estimates for each of these TE insertions were taken from Bergman and Bensasson (2007) based on the unique substitution method.

TE insertion alleles were assayed by PCR in 12 inbred wildtype strains of *D. melanogaster* from Zimbawe (GLINKA *et al.* 2003): A131, A145, A191, A398, A337, A229, A186, A384, A95, A157, A82, A84; and 12 inbred wildtype strains of *D. melanogaster* from North Carolina, USA selected randomly from the *Drosophila* Genetic Reference Panel (MACKAY *et al.* 2012): Bloomington *Drosophila* Stock Center IDs 25745, 25744, 25208, 25207, 25203, 25188, 25199, 25196, 25204, 25198, 25200, 25201. Genomic DNA from each strain was prepared using 30 adults. PCR cycling conditions were the same as described in Gonzalez *et al.* (2008) with some minor modifications for annealing temperatures. Two PCR reactions (to test for presence and absence of the TE, respectively) were conducted for each locus in each strain and the presence/absence of TE insertions was scored according to the same criteria as in Gonzalez *et al.* (2008). Loci that exhibited both presence and absence bands in a given strain were scored as heterozygous (FBti0019430, FBti0019165, FBti0019602, FBti0020077) and two alleles were counted as being sampled at this strain instead of one (coded as POLYMORPHIC in File S1). PCRs that failed 3 times in a given strain were scored as missing data (coded as NA in File S1). The frequency of the TE insertion in the North American or African sample was estimated as the number of presence alleles over the total number of alleles sampled (corrected for heterozygous loci and missing data). Summaries of the numbers of alleles sampled, observed allele frequencies, age estimates and other metadata for each locus can be found in File S2. We note that these PCR data have been independently shown to have greater than 92% concordance with *in silico* TE insertion predictions based on whole genome shotgun sequences from the same strains of *D. melanogaster* (CRIDLAND et al. 2013).

*Determination of probability distributions for TE insertion alleles under different models of host demography.*

To account for non-equilibrium host demographic history in our analysis, we allowed population sizes to vary over time based on published demographic scenarios for African and North American populations (LI and STEPHAN 2006; DUCHEN et al. 2013). For all calculations, the mutation rate of $1.45 \times 10^{-9}$/bp/generation from Li and Stephan (2006) was used to facilitate the use of these previously estimated demographic scenarios. For African samples, demography was also modeled according to Li and Stephan (2006). This assumes a current effective population size $N_e = 8.603 \times 10^6$ and time was scaled to correspond to a five-fold expansion (to current effective population size) that occurred 600,000 generations ago. For the case of the African samples, consideration must be made to the fact that the reference genome sequence used to ascertain the insertions was of North American origin and young insertion alleles present in the reference genome are thus unlikely to be sampled in Africa. For this reason, the analysis of the African sample should be seen mostly as a comparison to illuminate the behavior of the model under a different demographic scenario.

For the North American populations, a demographic scenario was modeled that approximated previous estimates (LI and STEPHAN 2006; DUCHEN et al. 2013). Under this scenario, the North American population is derived from a European population, which itself is derived from the African population. In particular, considering a current effective population of $1\times10^7$, time was scaled to correspond to a 300-generation bottleneck of 10,000 individuals that occurred 1,100 generations ago (Europe to North America migration), a European population of $1.075\times10^6$ individuals, and a 3,400 generation bottleneck of 2,200 individuals that occurred 154,600 generations ago (Africa to Europe migration). For the North American sample, we also consider a scenario of constant population size of $1\times10^6$ individuals. This latter scenario serves to correct potential biases that may arise from using a Bayesian posterior distribution for time since insertion when there are changes in population size (see below). Since the distribution of estimated TE ages has a large number of young TEs and also a long tail of older TEs (BERGMAN and BENSASSON 2007), we considered two parameters for the exponential prior distributions: $\lambda = 5.3\times10^{-6}$ and $1.875\times10^{-7}$. Both of these parameters were determined by fitting the TE age distributions empirically, either for only the youngest 143 elements or for the entire distribution. Probability distributions were calculated using both of these priors separately and final probabilities were determined as the weighted sum of the posterior probabilities, weighted by the relative likelihood of the number of observed substitutions for each element under these two priors. To allow for admixture in North America populations from Africa (CARACRISTI and SCHLOTTERER 2003; YUKILEVICH *et al.* 2010; VERSPOOR and HADDRILL 2011), we replaced the proportion of putative African alleles from the sample (determined by the expected level of admixture, assumed to lack the insertion) with the number of alleles that would be expected in this subsample under neutrality in the North American population.

All calculations were performed in *Mathematica 8* using numerical integration with 40 recursive bisections when needed. A Mathematica notebook to run the calculations presented here can be found in File S3. Results for the African population under an exponential prior and varying population size can be found in File S4. Results for the North American population under an exponential prior and varying population size can be found in File S5. Results for the North American population under an exponential prior and constant population size can be found in File S6.

*Forward simulations of transposable element dynamics*

To understand how our model performs under conditions where transpositional and demographic history are known, we performed two sets of simulation experiments under the extreme case of a single burst of transposition. This is a conservative test because we seek to determine the robustness of this method for testing neutrality when insertion alleles are not at equilibrium and display widely different frequencies. To model these dynamics, we considered the fate of a large number of TE insertion alleles whose frequency was simulated using a Wright-Fisher process. Since linkage disequilibrium is low in *D. melanogaster* (MACKAY *et al.* 2012), it is reasonable to assume that insertion alleles are independent.

In the first set (designated "time known") we simulated forward-time, neutral Wright-Fisher processes assuming a haploid population size of 1000 where a new TE insertion allele inserted at time zero with an initial frequency 1/1000. Since the majority of new neutral mutations are lost by drift, 10,000,000 replicate TE insertions were simulated to ensure that TE insertion alleles were retained in roughly 10,000 replicates. After a specified time in $N_e$ generations, the simulation was stopped. Individual TE insertions from replicate simulations were indexed and allelic state for each locus was randomly allocated to individuals to generate haploid genomes, each containing a set of unlinked TE insertions. From this population of 1000 haploid individuals, a single reference individual was selected and the allele frequency in a larger sample of 12 additional individuals was determined for each of the approximately 10,000 TE insertions ascertained from the reference. In some time known scenarios, negative selection was simulated by adjusting the relative sampling probability of a TE insertion during the Wright-Fisher process. Simulations with selection were performed only for recent transposition bursts since most deleterious elements become eliminated after a reasonable period of time has elapsed.

In the second set of simulations (designated "time unknown"), neutral Wright-Fisher forward simulations were performed as before, but instead of conditioning on a known number of generations, the number of substitutions within each insertion was simulated under a Poisson distribution. To approximate a population size on the order of one million, the $1.45 \times 10^{-9}$/bp/generation mutation rate was scaled 1000 fold and simulations were performed in a haploid population of 1000. In some time unknown scenarios, the population size was changed during the simulation. For each time unknown scenario, 190 TEs were selected from a randomly chosen single reference to model our actual dataset, and the probability of observing as many or fewer in the entire sample was determined for each of the 190 insertions, conditional on the number of simulated mutations and also the specified demographic scenario for the population size.

## Results

*Analysis of model predictions in a sample with known time of insertion.*

We developed a neutral population genetic model to test the evolutionary forces acting on TE insertions in natural populations that utilizes age information contained in the sequence of the TE itself. To illustrate the application of our model, Figure 2 shows the probability distribution of numbers of copies (from $i = 1$ to $i=n$) in a sample size of $n = 13$. To verify the accuracy of these predictions, we conducted forward simulations to generate sample frequencies for TE insertion alleles under the same scenario to account for ascertainment from a single reference genome (see Methods for details). Comparison of the results of these simulations to theoretical predictions under our model show strong agreement (Figure 2). Under the non-equilibrium transposition rate scenario simulated, completely neutral TE insertion alleles that occurred very recently in the past are expected to be at low frequency (Figure 2A). Conversely, TE insertions that have occurred distantly in the past are expected to be found in all sampled alleles since they will have coalesced prior to the insertion event, backwards in time (Figure 2D). At intermediate values of $t$ (measured in unit of $N$), the probability distribution of number of copies in a sample becomes nearly flat (Figure 2B). As previously noted by others (FELSENSTEIN 1992; SHERRY *et al.* 1997), if a mutation has occurred when all but two members of a sample of size $n$ have coalesced, it is equally likely that the mutation is represented in 1 to $n$-1 copies in the sample. Thus, an insertion of intermediate age will have a very flat probability distribution with high variance. For these reasons, there is little power to detect deviations from neutrality for single TE insertions at intermediate age. The power to detect general deviations from neutrality using our approach therefore lies in using TE insertions of varying ages to determine how well observed allele frequencies are correctly predicted by expected frequencies across many loci. We also conducted simulations under a scenario of negative selection acting on TE insertions that arose from a single recent burst of transposition. In this scenario, TE insertions of a given age segregate at lower frequencies than neutral insertion alleles as expected and show clear differences in frequency from model predictions (Figure 2A). Because selection eliminates most deleterious alleles quickly, we did not perform simulations of negative selection for older transposition bursts (2 B-D).

*Analysis of model predictions in a sample with estimated time of insertion.*
In the previous section, we verified that our model makes reasonable predictions when the time of insertion is known exactly. However, for insertions ascertained empirically from a reference genome, the time of insertion can only be estimated. To test the suitability of our model under more realistic assumptions, we used an empirical Bayes approach in which the posterior probability distribution of time since insertion is conditional on a simulated number of mutations and a prior distribution of possible insertion times. We considered two classes of priors in our model and tested their suitability using forward simulations. In one case, we used a uniform prior representing the span of ages estimated from the copy with the greatest number of substitutions. The uniform prior performed poorly and predicted insertion alleles to be at frequencies higher than observed (results not shown). We also evaluated the use of either an exponential or gamma distribution of times since insertion, with the exponential being a special case of the gamma. For recent bursts, where the mean number of substitutions per element is zero, the exponential is more appropriate. In simulations where the transpositional burst occurred at a sufficient time in the past, very few insertions will have accumulated zero substitutions. In these scenarios, a gamma-distributed prior is more appropriate. Simulations were performed again by allowing for a single transpositional burst within each population, but now the posterior time since insertion was estimated based on substitutions that were simulated by a Poisson process. For each of the transpositional bursts that occurred at a given time in the past, the parameters for the empirical prior (exponential or gamma) were estimated based on the Poisson distribution of substitutions. For constant population size simulations, we simulated four transpositional bursts at different times. We also consider the case of a transpositional burst that occurs at a time close to a rapid expansion in host population size.

To characterize the behavior of this approach, we simulated a sampling strategy similar to the one we actually used for the experimental data in this study. In particular, we simulated transpositional bursts in populations from which one reference individual was used to ascertain 190 TE insertions that were then sampled from 12 additional individuals. For each simulated population, we used our model to determine the distribution of 190 p-values for observing as many or fewer copies in the sample for each insertion allele under the neutral model. If our model is biased towards falsely rejecting the null hypothesis because it systematically predicts lower TE frequencies than expected under the neutral simulations, we would expect the distributions of p-values to be skewed toward zero.

These simulation results show that under constant population size, p-values for the probability of observing as many or fewer insertion alleles under our model are not biased. This is seen at each of the four transposition burst times (Figure 3A). For extremely recent transposition burst times (Figure 3A, $t=0.002$ and $0.01$) there is very little variation in the distribution of p-values. This is because nearly all such insertions have experienced insufficient time to either accumulate mutations or be found in any other individuals besides the reference. Thus, nearly all these insertions have the same p-value. Critically, even though these represent very low frequency insertion alleles, the null hypothesis of neutrality is not spuriously rejected.

While the behavior of our model is correct for constant host population sizes, simulations revealed that it can be biased under scenarios in which there are both

1  transpositional bursts and changes in population size (Figure 3B). For example,
2  we simulated populations that experience a transpositional burst and then, $t=0.2$
3  generations later (forward in time), experience a sudden ten-fold increase in
4  population size, followed by sampling at $t=0.1$ generations (scaled to the new
5  population size) later. We then tested whether frequencies were as predicted,
6  assuming a known demographic scenario but unknown age. In this case, the
7  distribution of p-values that as many or fewer insertion alleles are observed in
8  the sample are skewed toward zero (Figure 3B). Since times of insertion are not
9  precisely known, a significant part of the mass of the posterior distribution for
10 ages is greater than 0.3 (i.e. before the transposition burst during the period prior
11 to the population expansion). During this extended time, the population size is
12 much smaller and the rate of coalescence is faster, leading to an expectation of
13 higher allele frequency under neutrality relative to observed. As an illustration of
14 this effect, consider an extreme scenario in which a recent transpositional burst
15 occured two generations after a large and rapid increase in population size.
16 Under neutrality, the behavior of the insertions is predicted entirely by the new
17 population size. However, a Bayesian approach places a significant proportion of
18 the posterior probability for time of insertion in the era preceding the population
19 expansion. This leads to an increased expectation that the insertion alleles will be
20 at higher frequency. Under this scenario, our approach will therefore be correct
21 only to the extent that the posterior distribution of ages is similar to the actual
22 distribution of ages.
23
24 To account for this problem, one conservative approach to testing whether
25 negative selection is shaping allele frequencies is to model the current population
26 size only and ignore historical smaller population sizes. In our simulations, we
27 employed this approach by estimating the probability distribution using only the
28 current population size (Figure 3B, Constant Model). As can be seen, this
29 approach does not lead to spurious rejection of the null hypothesis and in fact is
30 highly conservative in a test for negative selection.

*Testing neutrality of TE insertions in D. melanogaster under non-equilibrium conditions.*

The age distribution of the 190 TE insertions sampled in this study indicates a large number of copies that have experienced either zero or few substitutions as well as a significant number that are much older (Figure 4). To fit this distribution of ages, we considered two different parameters for the prior exponential distribution. We consider a prior lambda for the exponential based on the mean number of substitutions for all TEs and also consider a separate lambda estimated for the very young TEs. Final probability distributions were weighted in proportion to the respective probabilities for observing the specified number of mutations under each of these two priors.

Using this general modeling framework, and keeping in mind the conditions under which this approach may be biased (see above), we applied our model to 190 TE loci in two populations of *D. melanogaster* (Figure 5). In the case of the North American sample, we determined how well the expected values under our model fit the data as a function of rank age of insertion estimates under a scenario of varying population size that included a substantial bottleneck in the migration out of Africa and also out of Europe (Figure 5A). Several observations are evident. First, Pearson's r for the overall correlation between observed and expected under the model is 0.85, indicating that the incorporation of age information can explain a significant amount of variation in insertion frequencies under neutrality. Second, the model predicts consistently higher than observed allele frequencies for young (insertions with zero unique substitutions) and middle-aged insertions (those with at least 1 substitution, up to 0.9% divergence). At face value, this result provides evidence for selection acting against TE insertion alleles limiting their increase. However, given our simulation results that modeling recent population growth can lead to overestimates in expected allele frequency, and given that the North American population of *D.*

*melanogaster* is known to have undergone recent population expansion (LI and STEPHAN 2006; DUCHEN *et al.* 2013), we suggest this result should be interpreted with caution (see below).

In contrast to the North American sample, fewer young alleles are segregating at intermediate frequencies in African sample. This is also expected in the African population because alleles take a longer time to drift to higher frequency in larger populations. It is also expected since the insertion alleles were ascertained from a non-African genome. Due to the larger population size and screening bias, more insertions are expected to be segregating at lower frequencies in Africa in contrast to North America. The results are consistent with this prediction. For the most part, TE insertions appear to either be segregating at either low or high frequencies in the African sample. Nevertheless, as with the North American population, the correlation between observed and expected allele frequencies under the model is quite high (Pearson's r=0.94). As such, the signal for negative selection acting against TE insertions in the African sample is not as strong as it is in the North American sample, although it is also evident for some moderately aged TEs in the Africa sample.

Many previous studies have shown an accumulation of TEs in regions of low recombination of the *D. melanogaster* genome (RIZZON *et al.* 2002; BACHTROG 2003; DOLGIN and CHARLESWORTH 2008). Our PCR data are consistent with this observation and our model also performs well in predicting the fixation of the older class of TE insertions largely residing in regions of low recombination (Figure 5). Likewise, previous work has shown that LTR elements are on average younger than non-LTR elements in *D. melanogaster* (BERGMAN and BENSASSON 2007). Consistent with this previous finding, observed allele frequencies for non-LTR insertions are typically higher than for LTR insertions in our sample (see also KOFLER *et al.* 2012). Jointly, low recombination rate regions of the genome (pericentromeric regions and chromosome 4) show a greater density of older non-LTR insertions that are mostly fixed. However, a lack of fixation can be observed for some LTR elements in low recombination regions of the genome, where they would otherwise be expected to be fixed (BARTOLOME and MASIDE 2004). These observations further support the idea that LTR elements are young in *D. melanogaster* and that young TE insertions will be at low allele frequency in this species.

*Accounting for bias when testing for negative selection on TE insertions*

As shown in Figure 5A, by conditioning on TE age and taking into account changes in population size, we observe that TE insertion alleles in North America are segregating at frequencies that are lower than expected. This suggests that negative selection is limiting the spread of TEs, and is consistent with the results of previous analyses that assumed constant transposition rates (CHARLESWORTH and LANGLEY 1989; PETROV *et al.* 2003; LOCKTON *et al.* 2008; GONZALEZ *et al.* 2009; LEE and LANGLEY 2010). However, this inference is confounded by several forms of bias that arise from the interplay between non-equilibrium host demographic history and uncertainty in the estimate of the age of TE insertions. As discussed above, when a transposition bursts occur close in time to a change in population size, using a Bayesian approach to estimating time since insertion can cause our model to predict frequencies higher than should be expected and lead to biases in inference. Additionally, our analysis of TE dynamics in the North American population in Figure 5 assumes a demographic scenario that does not account for admixture between North American and African populations (CARACRISTI and SCHLOTTERER 2003; YUKILEVICH *et al.* 2010; VERSPOOR and HADDRILL 2011).

To account for these issues, we took the conservative approach suggested by our simulation results (Fig 3B) by modeling the population size to be constant and equal to one million individuals. One million is slightly lower than the long term estimated effective population size of Africa (1,150,000: (CHARLESWORTH 2009)) and the current European population (1,075,000: (LI and STEPHAN 2006)). Under this scenario, the predicted effect of ancestral bottlenecks on allele frequencies is ignored. In addition, we also attempted to account for known admixture between North American and African populations that is estimated to be around 15% (DUCHEN et al. 2013). Specifically, we accounted for the possible effects of immigration of alleles from Africa that lack the TE insertion in lowering the observed TE frequency in North America by replacing 15% of the absence alleles at a locus with the expected number that would be derived from a sample of neutral alleles in North America.

Figure 6A plots the observed and expected North American frequencies under this revised scenario for the demographic history in North America. As anticipated, the observed and expected counts are more similar, since past bottlenecks are not influencing the predicted frequencies. Under this revised demographic model, Pearson's r for the overall correlation between observed and expected frequencies is 0.93, indicating a neutral model that is conservative can explain nearly all the variation in insertion frequencies. Under this conservative test, we find little support for the conclusion that selection acts to limit frequencies of the youngest TEs in our sample (i.e. those with zero substitutions). Many of these TEs may have inserted quite recently and are not expected to be at high frequency. Furthermore, since these TEs have zero substitutions, there is little power to distinguish their age from either having just transposed in the last few generations or further back in time, but not long enough ago to have accumulated a substitution.

In contrast, we do still observe lower allele frequencies for middle-aged TEs than expected under neutrality. As noted above, alleles of intermediate age are expected to be found at wide range of sample frequencies and for these alleles we do not have strong power to reject deviation from neutrality on an individual element basis. In aggregate, however, we find that for middle-aged insertions, the probability of observing as many or fewer copies is systematically skewed toward probabilities that are lower, with 23 p-values above and 62 p-values below 0.5 (Sign test: p<0.0001) (Figure 6B). Thus, even when we perform a conservative test of neutrality that accounts for potential bias in our method and admixture in the North American population, we still find evidence for negative selection acting to limit the frequency of middle-aged TE insertions in *D. melanogaster*.

In addition to these forms of bias due to non-equilibrium host demography, there are two possible sources of error by which TE insertion age (and therefore expected frequency) might be overestimated in our data and lead to a false signature of negative selection. One potential source of error would be caused by mutations that occur during the transposition process itself. For example, an error during the reverse transcription reaction would lead to a unique point substitution that would be incorrectly inferred have arisen after, rather than during, insertion. Studies of the *Ty1* retrotransposon in yeast indicate that this rate can be as high as $2.5 \times 10^{-5}$ per base pair (GABRIEL *et al.* 1996). We identified 154 TE insertions that were either young or middle-aged and showed evidence for negative selection. At an average size of 3789 base pairs, this would mean we expect about 15 of these 154 insertions have experienced such a mutation event during integration, assuming the rates for *Ty1* hold for the different TE families in our sample. To account for the effect of this potential source of error, we removed the 15 young or middle-aged TE insertions with the lowest estimated probabilities of being at their observed frequency.

A second potential source of error that would lead to over-estimation of TE age is if a all but one copy of an active sub-lineage in a family are lost or absent from the set of paralogous TEs sampled in the reference genome sequence. In this case, the age for the remaining insertion on that sub-lineage in the family would be over-estimated. To eliminate this problem, we identified ten middle age TE insertions that demonstrated a bias toward substitutions in the third position indicative of selective constraint on an active lineage. For these TEs, it is plausible that other representatives of the same sub-lineage may be absent from the reference genome sequence, leading to overestimation of time of insertion. After removing these ten TEs, the number of third position substitutions in the remaining set was identical to the average of 1st and 2nd position substitutions. After eliminating both classes of TEs whose ages are plausibly over-estimated (25 in total) as well as all putative adaptive TEs (see below) from the middle-aged set, we still observe a significant skew of p-values for middle age TEs in the North American sample, with 15 p-values above and 39 p-values below 0.5 (Sign test: p=0.0015). Thus even after applying demographic and age estimation corrections, we still find evidence that negative selection acts against middle-aged TEs in North American populations of *D. melanogaster*, despite low power to detect deviations from neutrality for this age class.

*Identification of candidate adaptive TE insertions.*

Despite general evidence for negative selection on many TE insertions, we also found evidence that several TE insertions are at higher frequency than expected and could therefore represent adaptive TE insertions. Under the constant population size model in the North American population, we find that the previously characterized adaptive Fbti0019430 *Doc* insertion in the *CHKov1* gene (PETROV *et al.* 2003; AMINETZACH *et al.* 2005) has a 0.19 probability of being as or more frequent in the sample. Using this probability as a liberal inclusive threshold (in light of the reduced power that occurs when we relax equilibrium assumptions), we identify seven other insertions that show higher frequencies than expected in North America in high recombination regions (Table 1). Within the African sample, we find two TE insertions that meet this criterion. One of these is a *Doc* insertion (FBti0019199) in the intergenic region between the genes *Pde11* and *CG15160* that is also found at higher than expected frequency in the North American sample, suggesting it is globally adaptive. Another candidate, a *412* element (FBti0020082) inserted between the genes *Or67a* and *Ir67a*, resides in a region that has previously been reported to show signatures of adaptive evolution (CONCEICAO and AGUADE 2010). Importantly, since this method conditions on age, it is capable of identifying alleles that are potentially adaptive but not fixed. For example, a *BS* element (FBti0020125) in the intron of the gene *CG43373* is present in only four of 12 African alleles sampled, but the probability of achieving such a high frequency under neutrality is 0.06. It should be noted that since the critical p-values for detecting putative adaptive insertions were made assuming a constant population size, they may be biased. An examination of the p-values under the model of varying population size (Table 1) indicates that many of these candidate adaptive TEs may have achieved the observed frequency by drift alone. Evidence for adaptation is strongest for insertions that are high in both Africa and North America (FBti0020125 and FBti0019199). Nonetheless, additional study is clearly required before concluding the insertions besides Fbti0019430 listed in Table 1 are adaptive.

**Discussion**

Here we show that the number of substitutions that have occurred on a TE sequence after its insertion in the genome can be used to test the neutrality of the allele frequency of that TE in a population sample. In so doing, we remove the need to assume anything about the transposition rate of TEs (at either the copy or family level), and as a consequence relax the assumption of a fixed transposition rate that underpins most models of TE evolution such as transposition-selection balance. Our model is also able to account for aspects of host demography that may confound the interpretation that TE insertion alleles have been driven to high frequency by selection rather than drift. Application of our model to a North American and an African sample of *D. melanogaster* shows that the age of a TE allele can explain more than 80% of the variation in allele frequency under complete neutrality. This demonstrates that it is important to take age structure of TE insertions into account when testing models of TE evolution. We also provide evidence to confirm the prevailing view that many TE insertions are likely under negative selection in a North American population of *D. melanogaster*, even though they may have been proliferating by periodic bursts of activity in this species (BLUMENSTIEL *et al.* 2002; BERGMAN and BENSASSON 2007).

Furthermore, using this method we were able to identify a small number of putatively adaptive TE insertions, including one (Fbti0019430) that was previously identified to be a target of positive selection (PETROV *et al.* 2003; AMINETZACH *et al.* 2005). However, when cross-referenced with two other studies that identified potentially adaptive TEs by different methods (GONZALEZ *et al.* 2008; KOFLER *et al.* 2012), only Fbti0019430 was found as a candidate in all three studies. This suggests that inferences of positive selection on TEs may be model dependent and that a joint approach using all three methods will be useful in screening for all possible sites of adaptation due to TE insertion. Further work, such as examining patterns of nucleotide variation in regions flanking TE insertions for signatures of selective sweeps (AMINETZACH *et al.* 2005; GONZALEZ *et al.* 2008; KOFLER *et al.* 2012) and functional studies, will be necessary to show that these TE insertion alleles are indeed found in positively selected regions of the genome and to determine if the TE insertion is in fact the target of selection.

There are several caveats with respect to the method presented here for testing departures from neutrality of TE insertion alleles. The power of our approach depends jointly on the effective population size and the mutation rate of the species in question. *D. melanogaster* has an effective population size of the order of one million and a mutation rate of $1.45 \times 10^{-9}$ mutations/bp/generation. Thus, for an unconstrained 5 kb TE insertion, approximately thirty nucleotide mutations are expected during the sojourn time between insertion and fixation, and we should have reasonable power to detect deviations from neutrality in this species. For substantially smaller populations, the time scale of mutation will be less than the time scale of drift to fixation within the population and there will be less power to detect deviations from neutrality with this method.

In addition, this method assumes there are not strong systematic errors in age estimation of TE insertions. Such errors could arise either from poor genome assembly of repeat sequences, inaccurate estimation of terminal branch lengths, or gene conversion events across dispersed repeat sequences that erase age information. It is unlikely that assembly quality impacts our results since TEs in *D. melanogaster* have been finished to high quality (CELNIKER *et al.* 2002; KAMINKER *et al.* 2002). Likewise, at least for the LTR retrotransposons used here, age estimates based on terminal branch lengths are likely to be reasonably accurate since they correlate with independent age estimates based on intra-element LTR-LTR comparisons (BERGMAN and BENSASSON 2007). If gene conversion among paralogous TE in indeed ongoing in the *D. melanogaster* genome, this source of error does not appear systematic because it would lead to global underestimation of true insertion age, which in turn would incorrectly lead to a prediction of lower insertion frequencies than is actually observed. For the demographic scenario that is most strongly supported by the population genetic data presented here, allele frequencies were in fact predicted to be higher than observed, opposite to the effect expected under pervasive gene conversion among paralogs. However, this issue is of concern for TEs that we classify as potentially adaptive, since these TEs that have experienced homogenization by gene conversion might in fact be older than their estimated age and therefore segregating at a high frequency as expected under neutrality.

Additional caveats relate to the use of a Bayesian approach to estimate the age of TE insertions when dealing with very young TEs and when transposition bursts occur close in time to host population expansions. Many young, zero-substitution TE insertion alleles were in fact not found in any strains in the population sample besides the reference genome. The interpretation that negative selection is acting to prevent these young TEs from reaching modest frequency implicitly depends on the assumption that these zero-substitution TEs represent a range of ages or that other slightly older TEs within the zero-substitution class have been removed from the population by selection and are therefore not to be found in the reference genome. In this regard, our method still shares some affinity with methods that make assumptions of transposition-selection balance (CHARLESWORTH and LANGLEY 1989; PETROV *et al*. 2003), since it generates an expected frequency in the population based on a theoretical distribution of insertion ages, not precisely known ages. Bayesian estimation of TE insertion age also can lead our model to generate incorrect predictions about allele frequency when bursts of transposition occur close in time to changes in host population size. In such cases, a significant part of the mass of the posterior distribution for estimated allele ages can be placed before or after the actual time of insertion, leading predictions of the model to be influenced by population sizes not experienced by the insertion. Thus, when testing neutrality, it is important to condition on a demographic scenario that is conservative with respect to the manner in which neutrality may be rejected.

Despite these caveats, our work provides an advance over previous work in several regards. We show that an age-based test of neutrality can be constructed that takes advantage of the molecular evolutionary information intrinsic to large insertion mutations like TEs. This result permits development of a new class of models to test the general mode of evolution of TE insertions that relax the assumption a fixed transposition rate, an assumption that is highly unlikely given what is known about the biology of TEs but which currently underlies models of transposition-selection balance. Such a test may be beneficial in determining how selection against TEs varies among species, because it can take into account differences in the histories of TE proliferation. In addition, this method is capable of eliminating, without a defined age threshold, the older class of TE insertions as being candidates for recent adaptation. It also discriminates against detecting high frequency insertions that may appear to be young, but in fact lack substantial age information. For example, one *G4* element (FBti0019755, Rank #17 in Figure 5) is found at high frequency but has zero substitutions. However, the age estimate for this insertion is based on only 40 bp of sequence and is therefore unreliable, and thus this TE fails to meet the threshold of being at an unusual frequency given its age.

Importantly, TEs are not the only form of insertion mutation that have this additional age information, and thus our approach could be extended and applied to other insertion alleles, such as gene duplications and other copy number variants. If the number of substitutions that have occurred since duplication can be estimated (for example, from silent sites or intronic regions, assuming no purifying selection is acting at these positions), one may also ask whether the allele frequency of new gene duplicates are consistent with neutrality using the approach developed here.

**Acknowledgements**

We thank Wolfgang Stephan and Nicolas Svetec for providing African strains of *D. melanogaster*; Dmitri Petrov for providing PCR primer sequences; Dan Hartl, John Wakeley, Brian Charlesworth, Adam Eyre-Walker, Daniel Živkovi•, Matthew Ronshaugen and Maria Orive for helpful discussion during the project; and Stephen Wright and two anonymous reviewers for helpful comments on the manuscript.
34

**Figure Legends**

**Figure 1**. Method for estimating TE insertion age based on unique substitution counts from insertions gathered from a single reference. A) i) Schematic of evolutionary dynamics for two active sub-lineages of the same TE family, depicting recent transposition events (arrows) leading to new TE insertions (rectangles) and post-insertion mutation events (black tick marks inside rectangles). Each horizontal line represents a single chromosomal segment in a population sample. Dashed lines indicate where segments lack a TE sequence relative to the reference genome. TEs located above segments are insertions not present in the reference genome. In this example, TE insertion **a** has recently integrated, is at low frequency in the population sample and has accumulated no unique mutations. In contrast, TE insertions **b, c** and **d** represent older insertions that are at higher frequency in the population which have accumulated unique mutations. ii) Schematic depicting the procedure used to estimate the age of TE insertions identified in the reference genome. A multiple alignment of all paralogous copies of the TE family from the reference is generated. Variant sites are identified and classified as being shared or unique, with only the number of substitutions unique to each reference insertion, $s$, being used to estimate the age since insertion. Shared substitutions are inferred to arise on active lineages and excluded from the estimate of allele age. Our model contrasts age based on $s$ with TE insertion allele frequency in the population, $i$. Older reference insertions with higher $s$ are expected to have a greater frequency $i$ under neutrality. B) Schematic of coalescent process for a TE insertion that is ascertained from a reference genome sequence. Frequency in the sample is a function of the number of descendants from a single ancestor that received the insertion at time $t$ and gave rise the reference insertion allele. In this example, insertion **c** from panel A inserted at the time at which the $n=7$ sample alleles have $j=3$ ancestors. All descendants from the insertion contain the insertion allele ($i=3$). Since the time of insertion, $s=2$ unique substitutions have accumulated on the reference insertion.

It is only these unique substitutions leading to the reference allele that are used to estimate the age of the TE insertion. Other mutations arise independently on non-reference insertion alleles, which could in principle be used to estimate the time to the most recent common ancestor (TMRCA) of the insertions allele, but are *not* used here.

**Figure 2**. Probability for *i*, number of insertion copies in the sample, under model predictions and simulations. *t* indicates known time since insertion. Selection was only simulated for the case where t=0.1 (A) because deleterious elements become quickly eliminated from the population at later times.

**Figure 3**. Distribution of p-values for observing as many or fewer insertion alleles, for 190 simulated insertion loci, where ages of each TE are estimated using the model from a Poisson simulated number of substitutions. Median p-value is indicated with a bold line, upper and lower quartiles with a box, range with whiskers and outliers with dots. A) Effects of time since insertion, *t*, on model based inference. A constant population size of $N_e$ = 1000 was simulated with varying time of insertion = *t*. Inference under the model used constant $N_e$. B) Effects of varying $N_e$ on model based inference. After a transposition burst, a population of 100 was simulated for 20 generations (*t*=0.2) followed by expansion to 1000 individuals for 100 generations (*t*=0.1) for a total *t*=0.3. Inference under the model was performed in two ways. Under the varying model, the probability of observing as many or fewer alleles was estimated, conditional on the same demographic scenario that was simulated. Under the constant model, the probability of observing as many or fewer alleles was estimated, conditional on a constant (post-expansion) population size of 1000.

**Figure 4.** Distribution of ages (in *s*, unique subs/bp) of the 190 TEs used for this analysis.

**Figure 5**. Observed and expected allele counts under models of varying population size for North American and African populations of *D. melanogaster*. In both panels, loci are ranked by age and the analysis accounts for age uncertainty and ascertainment bias. A) Observed and expected allele counts in the North American sample assuming the demographic scenario of a bottleneck from Africa to Europe followed by a bottleneck from Europe to North America. B) Observed and expected allele counts for the African demographic scenario of an ancient population expansion. See methods for details of demographic scenarios. Between panels, TEs from low recombination rate regions and non-LTR families are indicated.

**Figure 6.** A) Observed and expected allele counts assuming a constant population size for a North American population of *D. melanogaster*. In both panels, loci are ranked by age, the analysis accounts for age uncertainty and ascertainment bias and observed counts are also adjusted for admixture. B) Probability of observing as many or fewer copies in the sample for each TE.

# Tables

**Table 1.** Candidate adaptive TE insertions in North American and African populations of *D. melanogaster*. Shown are expected allele frequencies and the p-value of observing as many or more alleles in the sample under varying or constant demographic scenarios as described in the main text.

| | | | | North America | | | | | Africa | | |
|---|---|---|---|---|---|---|---|---|---|---|---|
| Flybase ID | Family/Order | Substitutions/Length | Recombination | Observed | Expected (Constant) | p-value (Constant) | Expected (Varying) | p-value (Varying) | Observed | Expected (Varying) | p-value (Varying) |
| FBti0019200 | Doc/non-LTR | 0/4336 | high | 3/13 | 1.5/13 | 0.11 | 4.32/13 | 0.48 | 1/13 | 1.05/13 | 1 |
| FBti0020082 | 412/LTR | 0/4972 | high | 5/10 | 1.3/10 | 0.01 | 3.13/10 | 0.26 | 1/6 | 1.02/6 | 1 |
| FBti0020086 | 17.6/LTR | 5/5814 | high | 6/13 | 1.8/13 | 0.02 | 5.87/13 | 0.49 | 1/13 | 1.09/13 | 1 |
| FBti0020149 | BS/non-LTR | 1/4579 | high | 12/13 | 1.96/13 | 0.00 | 6.25/13 | 0.15 | 1/13 | 1.11/13 | 1 |
| FBti0019354 | 17.6/LTR | 2/5787 | high | 5/13 | 2.32/13 | 0.10 | 7.42/13 | 0.71 | 1/13 | 1.15/13 | 1 |
| FBti0020046 | Doc/non-LTR | 1/2138 | high | 5/13 | 2.66/13 | 0.15 | 7.02/13 | 0.69 | 1/13 | 1.31/13 | 1 |
| FBti0019430 | Doc/non-LTR | 11/4323 | high | 13/13 | 8/13 | 0.19 | 9.34/13 | 0.33 | 5/15 | 5.27/15 | 0.51 |
| FBti0019199 | Doc/non-LTR | 1/2627 | high | 8/8 | 1.84/8 | 0.00 | 4.37/8 | 0.16 | 11/12 | 1.20/12 | 0.00 |
| FBti0020125 | BS/non-LTR | 5/4579 | high | 5/12 | 4/12 | 0.34 | 7.40/12 | 0.76 | 5/13 | 1.90/13 | 0.06 |

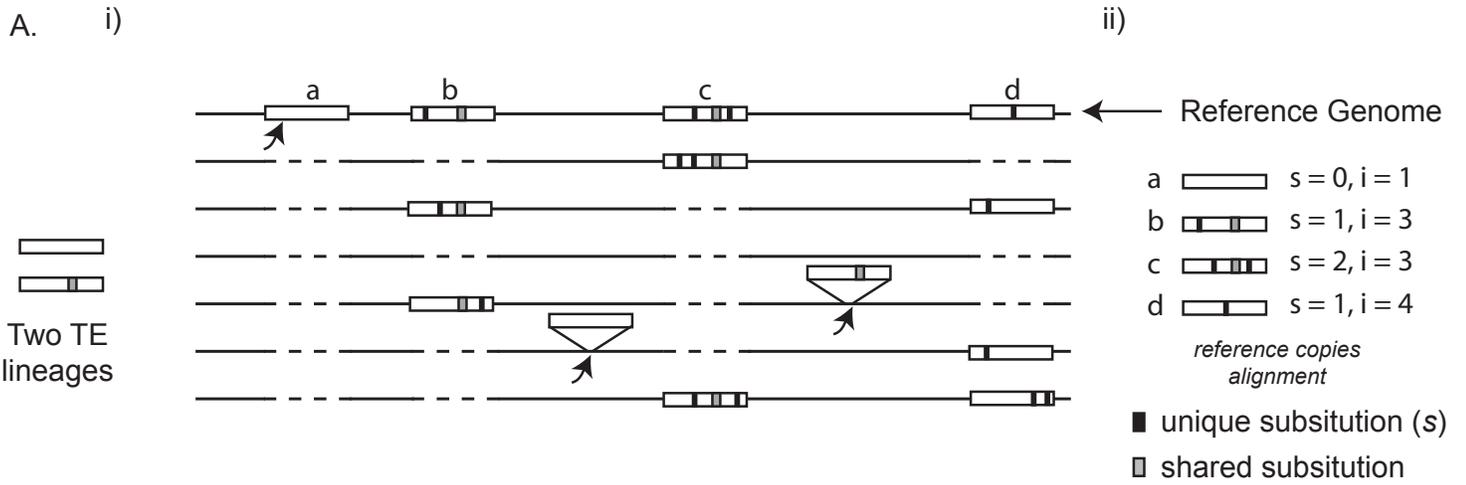
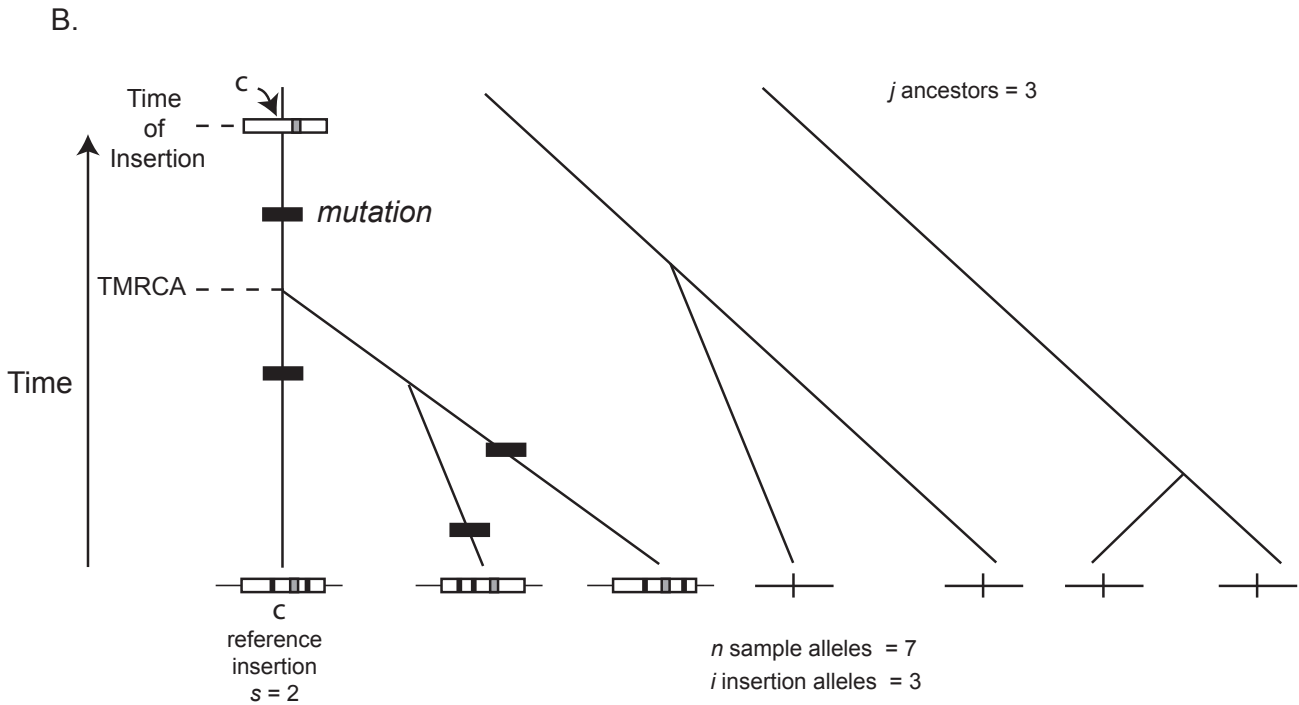

Figure 1.

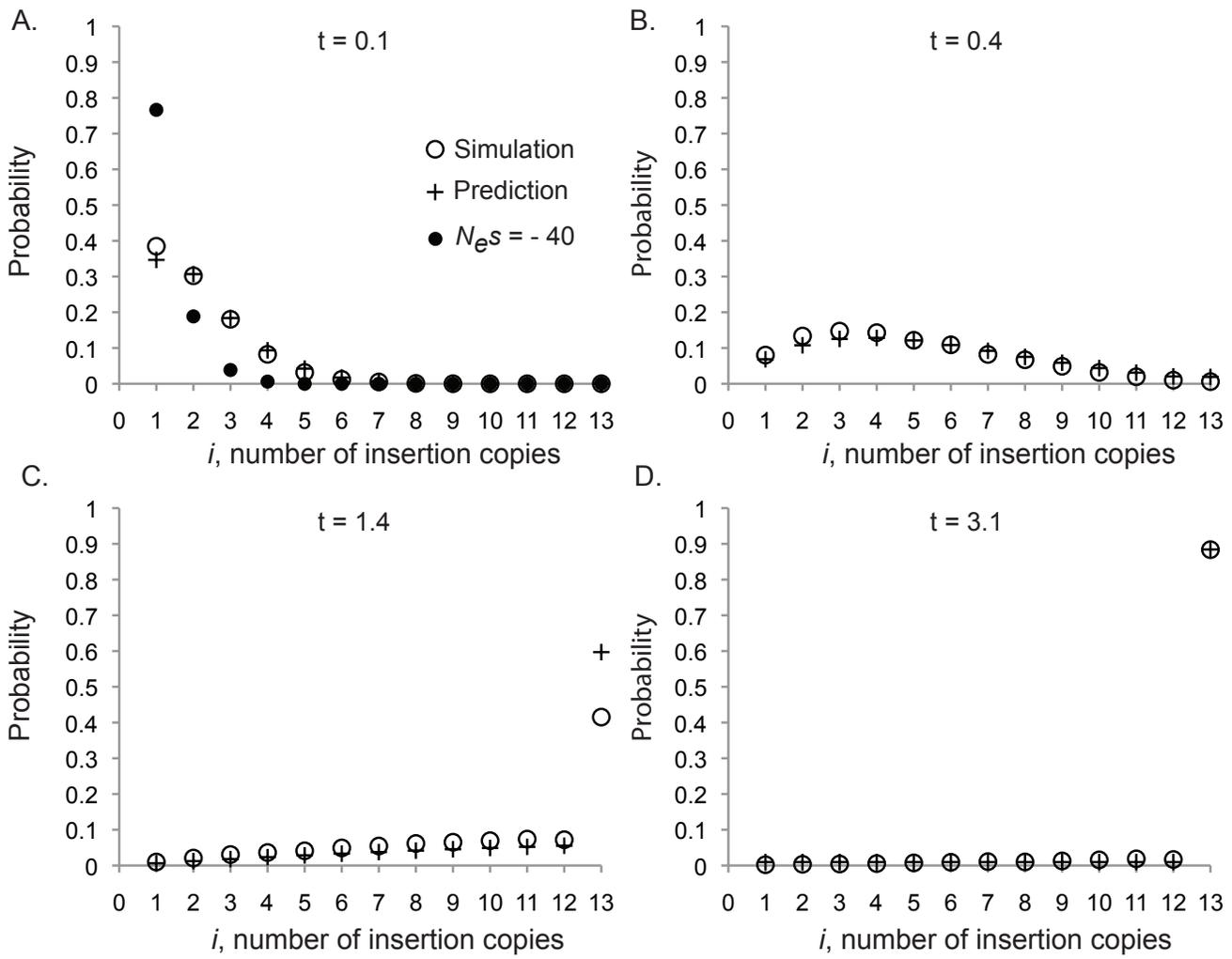

Figure 2.

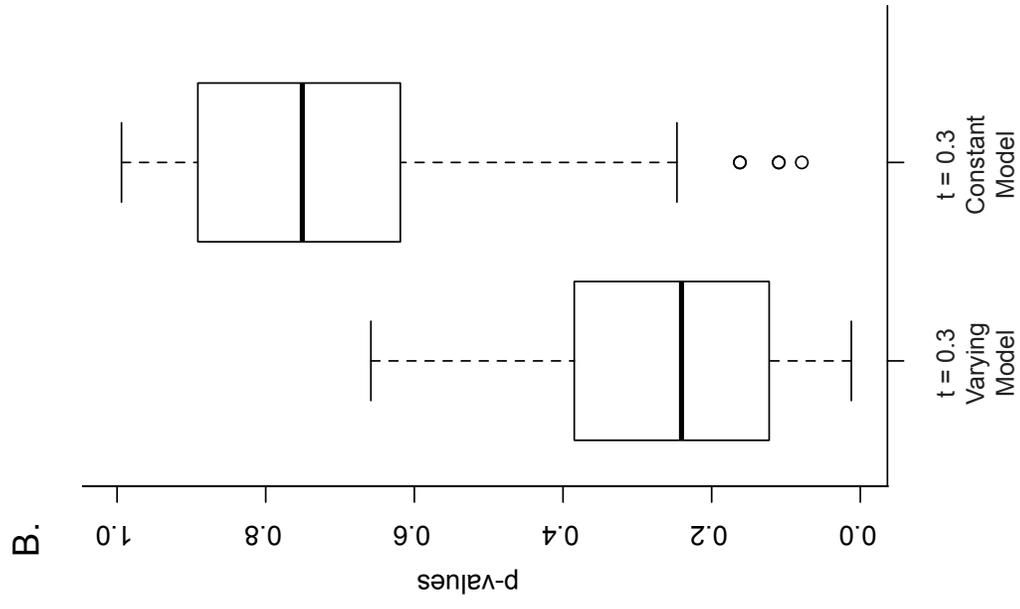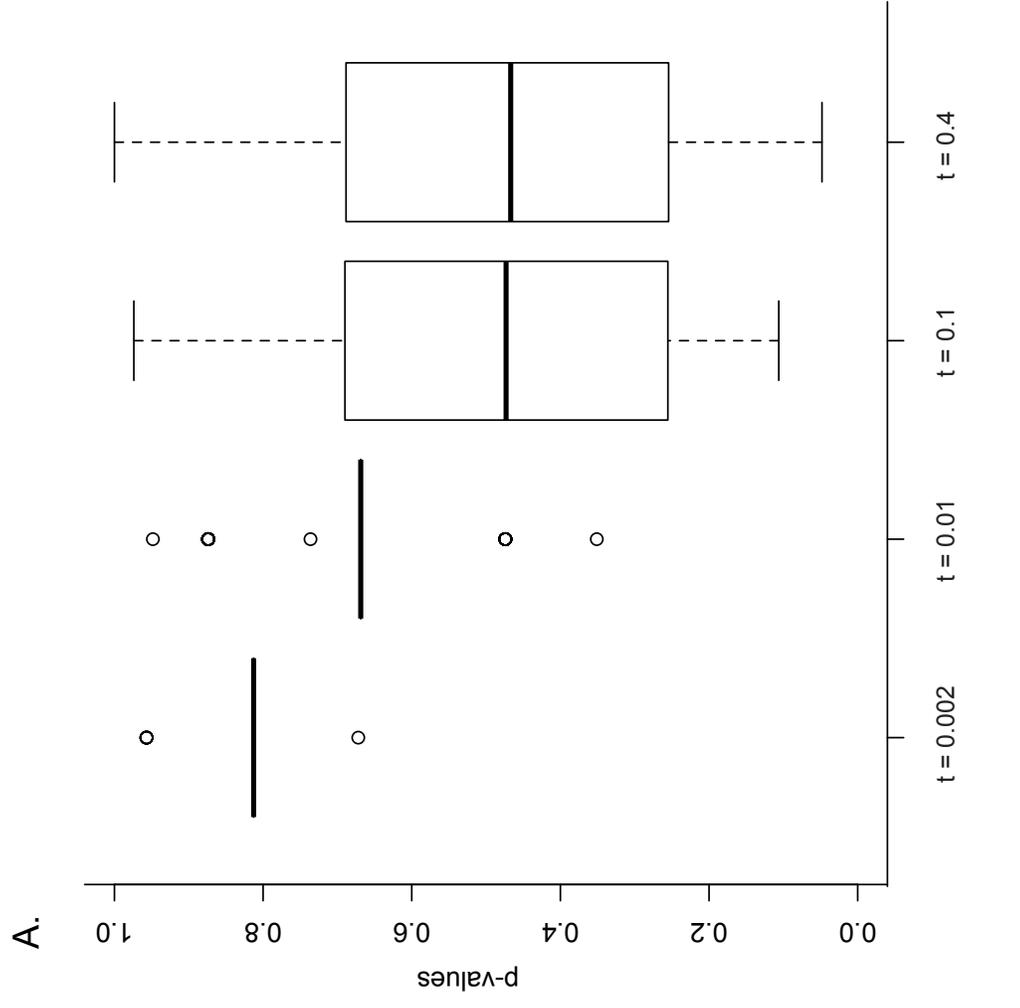

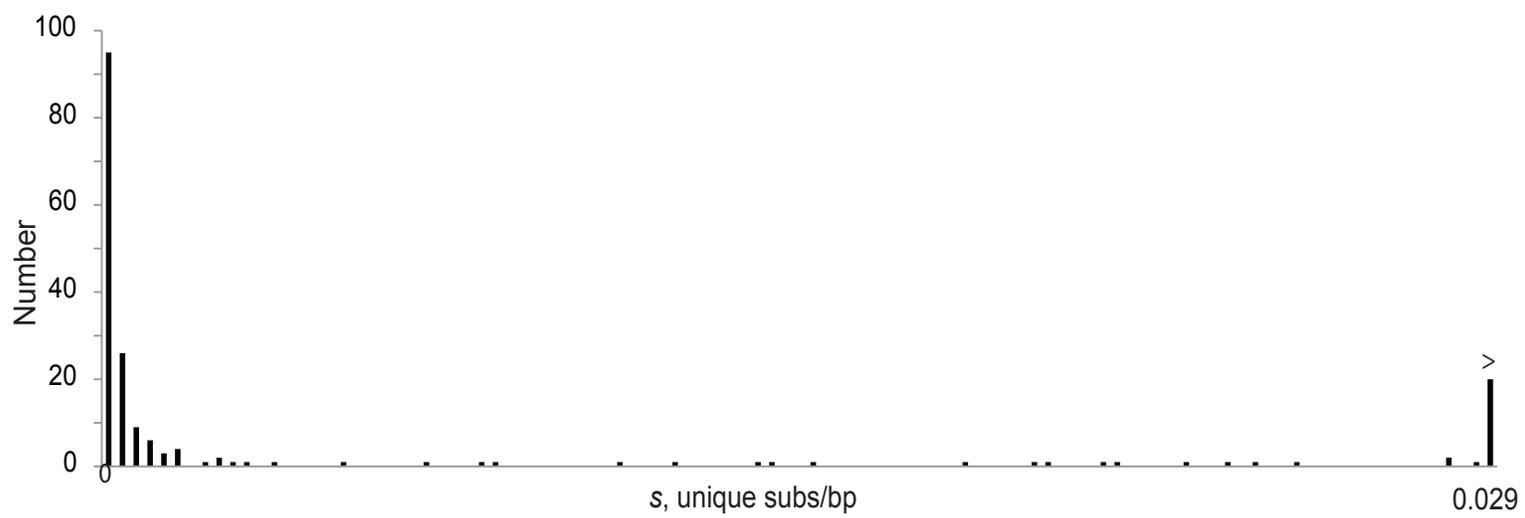

Figure 4.

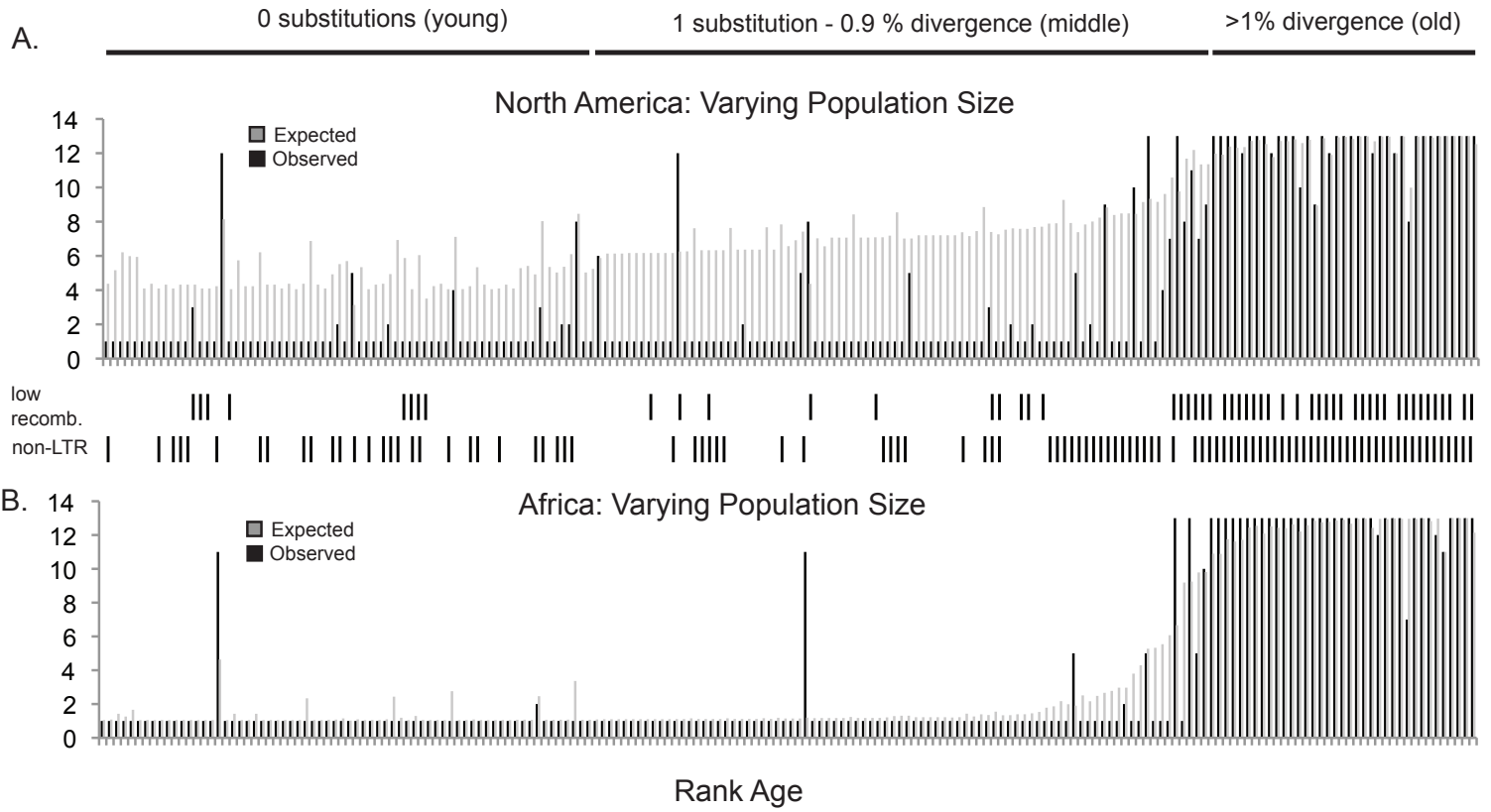

Figure 5.

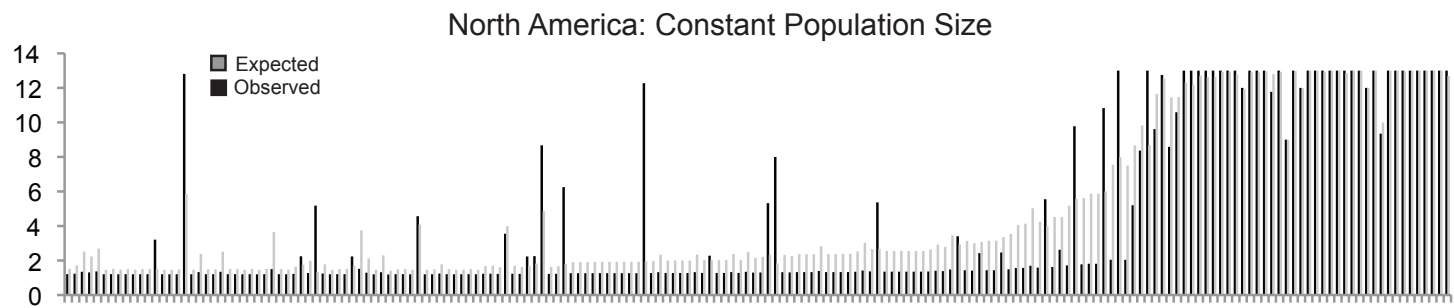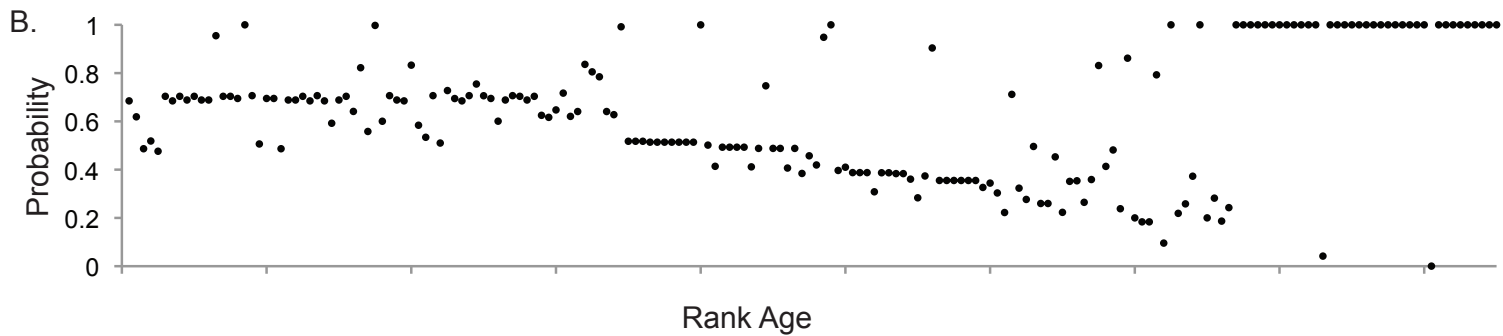